\documentclass[a4paper]{jpconf}
\usepackage{amsmath,amssymb}
\usepackage{graphicx}
\usepackage{color}
\usepackage[utf8]{inputenc}
\usepackage{dcolumn}
\usepackage{hyperref}
\usepackage{textcomp}
\usepackage{enumitem}

\usepackage{comment}
\usepackage[dvipsnames]{xcolor}
\usepackage{tikz}
\usetikzlibrary{calc}
\usetikzlibrary{decorations.pathmorphing}
\usetikzlibrary{decorations.pathreplacing}
\usetikzlibrary{decorations.markings}
\usetikzlibrary{shapes.misc}
\usetikzlibrary{positioning}
\usetikzlibrary{snakes}

\hypersetup{colorlinks=true,breaklinks,linkcolor=blue,urlcolor=blue,citecolor=blue}

\newcommand{\kv}{\ensuremath{\mathbf{k}}}
\newcommand{\qv}{\ensuremath{\mathbf{q}}}
\newcommand{\KV}{\ensuremath{\mathbf{q}}}
\newcommand{\av}[1]{\ensuremath{\left\langle #1 \right\rangle}}

\tikzset{->-/.style={decoration={
  markings,
  mark=at position #1 with {\arrow{>}}},postaction={decorate}}}
\tikzset{-<-/.style={decoration={
  markings,
  mark=at position #1 with {\arrow{<}}},postaction={decorate}}}

\begin{document}

\title{The extended Hubbard model with attractive interactions}

\author{E G C P van Loon and M I Katsnelson}
\address{Radboud University, Institute for Molecules and Materials, NL-6525 AJ Nijmegen, The Netherlands}

\ead{evloon@science.ru.nl}

\begin{abstract}
We discuss the phase diagram of the extended Hubbard model with both attractive and repulsive local and nonlocal interactions. The extended dynamical mean-field theory (EDMFT) and the dual boson method (DB) are compared. The latter contains additional nonlocal correlation effects that are not incorporated in EDMFT. We find that EDMFT and DB give almost identical results in the attractive $V$ regime, where phase separation occurs. This is quite a difference with the previously studied repulsive $V$ regime, where EDMFT and DB give very different phase boundaries for the checkerboard order phase, especially at small $U$.
\end{abstract}

The Hubbard model~\cite{Hubbard63,Gutzwiller63,Kanamori63,Hubbard64} is the prototypical example of a system with strongly correlated electrons. It features competition between the kinetic and potential energy that leads to a plethora of interesting physics. In the Hubbard model, the potential energy is given by a purely local interaction $U$. In recent years, extensions of the Hubbard model which also include nonlocal interactions have also gained attention, often in the contex of two-dimensional systems such as surface systems~\cite{Hansmann13}, graphene~\cite{Schuler13}, transition metal dichalcogenides~\cite{vanLoon17} and ultracold fermions in optical lattices~\cite{DuttaPRA13,DuttaRPP15,vanLoon15-2}.

The extended Hubbard model is usually considered with repulsive interactions $V>0$ and $U>0$. This choice is motivated by physical considerations, since the Coulomb interactions between electrons is repulsive. In this work, we will also look at attractive interactions. Physically, a Hubbard model with attractive interactions is realizable in ultracold fermion experiments~\cite{Esslinger10} and also in the condensed matter context~\cite{vanderMarel88,Micnas90,Strand14}.
Our motivation for studying attractive interactions here, however, comes from theoretical considerations. Strong nearest-neighbor repulsion leads to a charge-ordered state. The phase transition line from the Fermi liquid to this ordered phase has been determined in several computational approaches~\cite{Micnas84,Zhang89,Bursill93,Avella04,Avella06,Amaricci10,Ayral13,vanLoon14-2,Lhoutellier15,Stepanov16b,Terletska17,Ayral17,Kapcia17} and currently serves as the main way to compare these methods. In the attractive $V<0$ regime, a second transition occurs, now to a phase separation into a high and a low density state~\cite{Lhoutellier15,Fresard16}. Our purpose is to use this transition as a new benchmark for computational approaches to the extended Hubbard model.

After the success of Dynamical Mean-Field Theory (DMFT)~\cite{Metzner89,Georges96} in the Hubbard model, Extended Dynamical Mean-Field Theory (EDMFT)~\cite{Sengupta95,Si96,Kajueter96,Smith00,Chitra00,Chitra01} was developed to incorporate nonlocal interactions. Just like DMFT, EDMFT includes local correlation effects via an effective impurity model. The idea of EDMFT is to encode the nonlocal interaction in a retarded interaction on the impurity level. Recent analysis, however, has shown that the predictions for the charge order transition in EDMFT do not match more advanced theories very well~\cite{vanLoon14-2,Stepanov16-2,Terletska17,Ayral17}. This should not come as a great surprise, the charge susceptibility intrinsically has a rich momentum structure, even in weakly interacting systems, and encapsulating this structure in a local quantity is difficult. To address this issue, new theories have been developed that add corrections with momentum dependence to (E)DMFT, either diagrammatically or via a cluster approach.

For the square lattice Hubbard model at half-filling, cluster-DMFT and the dual boson approach have given very similar results for the transition line between the Fermi liquid and the checkerboard charge density wave~\cite{Terletska17}. EDMFT overestimates the $V$ needed to stabilize the charge-ordered phase. In EDMFT+GW, the Fock exchange is apparently crucial to get an accurate prediction of the phase boundary~\cite{Ayral17}. With this situation in mind, we now turn our attention to attractive local ($U<0$) and nonlocal interactions ($V<0$).

\section{Model}

The extended Hubbard model is given by the Hamiltonian 
\begin{align}
 H = - t\sum_{\substack{\sigma \\ \av{jk}}} c^\dagger_{k\sigma}c^{\phantom{\dagger}}_{j\sigma} 
 + \frac{1}{2} U\sum_{j} n_{j}n_{j} 
 + \frac{1}{2} V\sum_{\av{jk}} n_{k}n_{j}.
\end{align}
Here, $c^\dagger_{k\sigma}$ is the creation operator of an electron on site $k$ with spin $\sigma$, $n_{j}$ is the density on site $j$ and $\av{jk}$ denotes that $j$ and $k$ are nearest-neighbors.
The parameter $t$ is the hopping amplitude, and we use $t=1$ to set the energy scale. The potential energy is given by the local Hubbard interaction $U$ and the nonlocal Coulomb interaction $V$.
We consider this model on the square lattice at half-filling, i.e., one electron per site.

The square lattice is highly symmetric, and this has big implications for the phase transitions that occur. A well known example is the Ising model on the square lattice, where the ferromagnetic transition for $J>0$ and the antiferromagnetic transition for $J<0$ show essentially the same behavior. The bipartite nature of the square lattice is responsible for this. Similar things happen in the Hubbard model~\cite{julich2015Mielke}. In particular, in the Hubbard model (that is, with $V=0$) on the square lattice the transformation $U\rightarrow -U$ interchanges charge and spin excitations and keeps the overall physics the same. 

The nonlocal interaction $V$ breaks this symmetry since it explicitly couples to charge excitations and not to spin excitations. Note that the potential energy still has a symmetry related to the bipartite lattice: $V\rightarrow -V$ interchanges checkerboard order and phase separation, in analogy with the antiferromagnetism-ferromagnetism symmetry of the Ising model. So for $t=0$, there would be a symmetry under $V\rightarrow -V$. It is the combination of kinetic and potential energy that finally destroys the symmetries. Of course, the combination of kinetic and potential energy is exactly what makes the Hubbard model interesting.

\section{Method}

Our main method is the dual boson (DB)~\cite{Rubtsov12} approach. We use the same methodology as in Ref.~\cite{vanLoon14-2} and give a short overview here. The calculation is a two stage process. We start by performing a self-consistent EDMFT calculation, and then include the nonlocal correlation corrections of DB on top of the EDMFT results. 

The EDMFT self-consistency procedure consists of the iterative solution to find an effective single-site problem, the impurity model, that gives a local self-energy and polarization so that the local properties of the impurity model and the lattice model correspond.

The impurity model is given by a hybridization function $\Delta_\nu$ and a screened interaction $U_\omega$, and solved using the continuous time quantum Monte Carlo solver~\cite{Rubtsov05,CTQMCRMP} of Ref.~\cite{Hafermann13}. This impurity solver uses improved estimators~\cite{Hafermann12,Hafermann14} to accurately determine the impurity correlation functions. It is based on the ALPS~\cite{ALPS2} libraries. Computationally, this is the most difficult part of the calculations.

Then, we calculate a single-shot of DB, using the ladder approach for the dual polarization operator, $\tilde{\Pi}$. It is given by~\cite{vanLoon14-2}
\begin{align}
\tilde{\Pi}_{\KV\omega} =& \frac{T}{N}
\sum_{\kv\nu}\lambda_{\nu+\omega,-\omega}
\tilde{G}_{\kv\nu} \tilde{G}_{\kv+\KV\nu+\omega}
\Lambda_{\qv\nu\omega} \\
\Lambda_{\qv\nu\omega}=&\lambda_{\nu\omega}-\frac{T}{N}\sum_{\kv'\nu'}\Gamma_{\qv\nu\nu'\omega}\tilde{G}_{\kv'\nu'} \tilde{G}_{\kv'+\KV\nu'+\omega}\lambda_{\nu'\omega} \\
\Gamma_{\KV\nu\nu'\omega} =& \gamma_{\nu\nu'\omega}\! -\! \frac{T}{N} \sum_{\kv''\nu''} \gamma_{\nu\nu''\omega}
 \tilde{G}_{\kv''\nu''}\tilde{G}_{\kv''+\KV\nu''+\omega} \Gamma_{\qv\nu''\nu'\omega}.
\end{align}
where, in the current approximation, $\tilde{G}=G-g^{\text{impurity}}$ and $\lambda$ and $\gamma$ are the fermion-boson and fermion-fermion vertex of the impurity model, $\kv$ and $\qv$ are momenta and $\nu,\nu',\omega$ are Matsubara frequencies. The temperature $T$ and the total number of sites in the lattice $N$ act to normalize the frequency and momentum sums.
These are called ladder equations, expansion of the recursive definition of $\Gamma$ leads to a ladder with an increasing number of rungs $\gamma$, and rails $\tilde{G}$.

From the dual polarization, we determine the physical susceptibility $\chi_{\omega,q} = \av{nn}_{\omega,q}$ according to the DB formula for the susceptibility
\begin{align}
 X_{\omega,q}^{-1} = [\chi_\omega (1+\chi_\omega \tilde{\Pi}_{\omega,q}) ]^{-1} + \Lambda_\omega - V_q. \label{eq:pi}
\end{align}
This formula illustrates the difference between DB and EDMFT, since EDMFT corresponds to $\tilde{\Pi}=0$.
The combination $1+\chi_{\omega=0} \tilde{\Pi}_{\omega=0,q}$ can be interpreted as the dual correction to the EDMFT result.
We also note that the susceptibility in this DB scheme satisfies the charge conservation law~\cite{Hafermann14-2} at $q=0$, i.e. $\chi_{\omega\neq0,q=0}=0$. 

It is also possible to go further in the dual theory, on the one hand by including further diagrammatic corrections to the polarization and to the nonlocal self-energy, on the other hand by taking into account feedback of the nonlocal correlation effects onto the impurity model~\cite{Stepanov16}. Here, we choose this relatively simple scheme since it allows for a direct comparison between EDMFT and DB and because it has been applied successfully to the repulsive ($U>0$, $V>0$) situation already~\cite{vanLoon14-2}.

Charge-ordered phases are indicated by a divergence of the zero-frequency susceptibility at some specific momentum $q$. The checkerboard order that occurs at $V>0$ is visible at $q=(\pi,\pi)$. On the other hand, at $V<0$, phase separation takes place, which is visible in a diverging compressibility, i.e. the susceptibility at $q=0$. All our calculations are performed in the Fermi liquid phase where all susceptibilities stay finite, the phase boundary is determined by extrapolating $\chi^{-1}_{\omega=0,q}$ as a function of $V$ and finding the value of $V$ where it crosses zero.

We study the two dimensional square lattice extended Hubbard model at half-filling and at fixed temperature $\beta t=10$.
In all our calculations, the system is a Fermi liquid at $V=0$. At sufficiently large $U$ and sufficiently low temperature, the system will turn insulating~(For recent study of the metal-insulator transition in the square lattice Hubbard model, see \cite{Schafer15}), but we do not consider that regime here. We also restrict our attention to the charge sector.

\section{Results}

\begin{figure}
\centering
\includegraphics{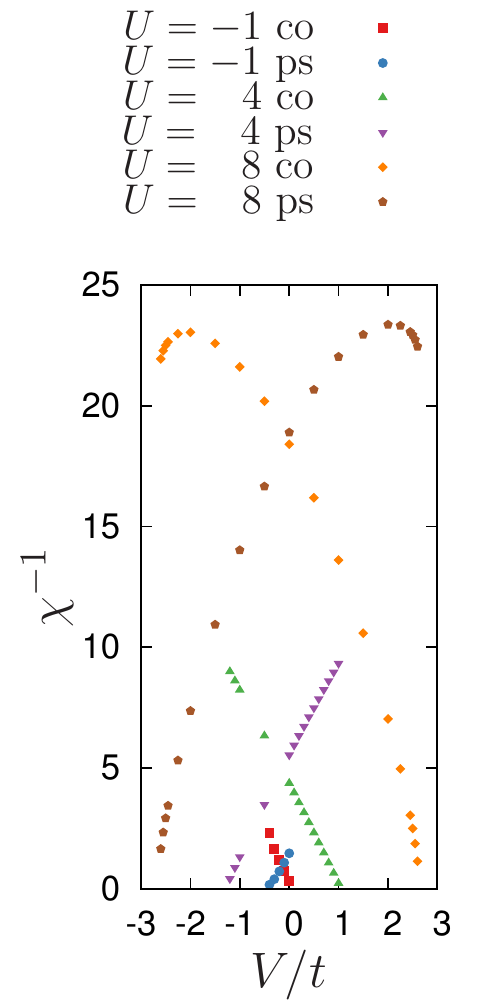}%
\includegraphics{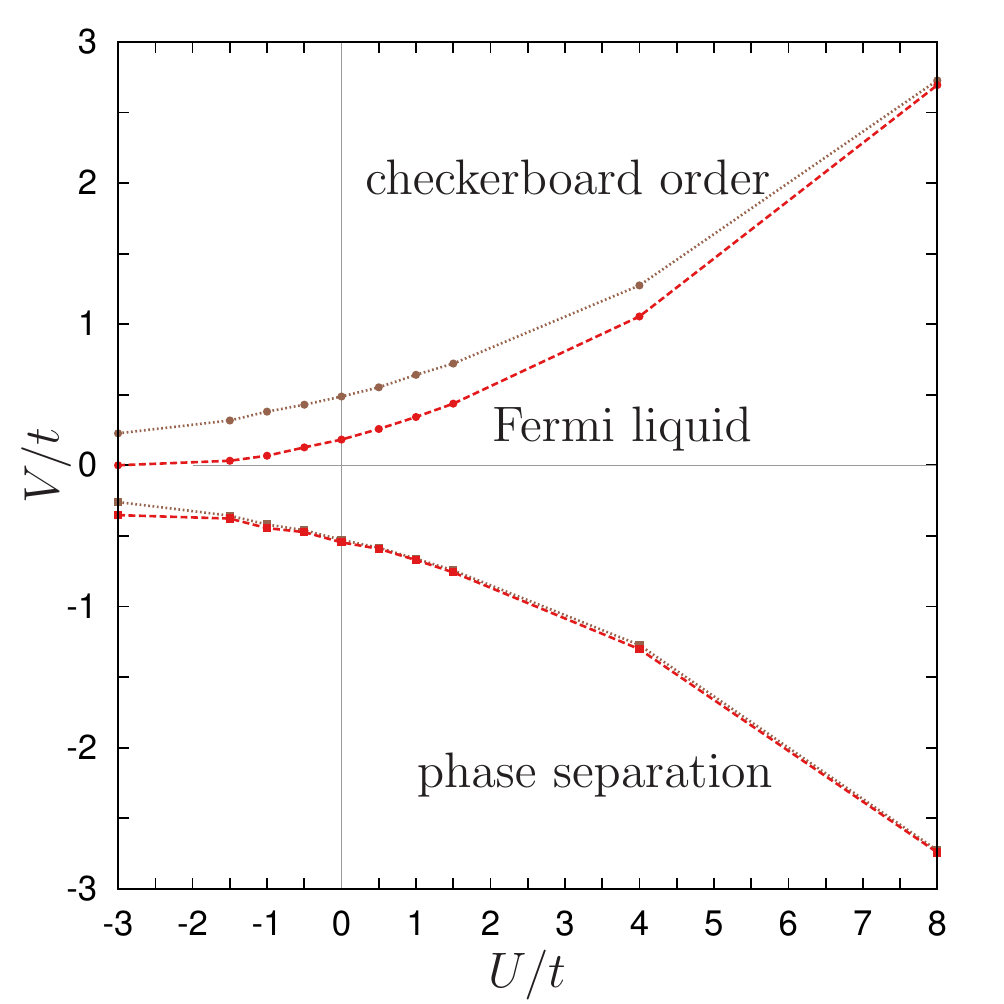}%
\caption{Left: Inverse susceptibility in DB as a function of $V$. The phase boundary is determined by extrapolating to $\chi^{-1}=0$. For every value of $U$, both the checkerboard order (co) and phase separation (ps) susceptibilities are shown, corresponding to $q=(0,0)$ and $q=(\pi,\pi)$. 
Right: Phase diagram at $\beta=10$. The red dashed lines are the DB results and the brown dotted line corresponds to the simpler EDMFT approximation. 
\label{fig:pd}
}
\end{figure}

In Fig.~\ref{fig:pd} we show the phase diagram of the extended Hubbard model according to the EDMFT and DB methods, at $\beta t = 10$. The left panel shows the inverse susceptibility. For fixed $U$, a larger value of $V$ increases the checkerboard susceptibility and usually decreases the uniform (phase separation) susceptibility. An exception to this behavior is found at $U=8$ and large values of $V$ close to the checkerboard order. Here, the uniform susceptibility actually increases as a function of $V$. A comparison at fixed $V$ shows that larger $U$ decreases the magnitude of the charge susceptibility. 

The phase boundaries in the right panel are obtained from the inverse susceptibility. The $U>0$, $V>0$ results are similar to those in the earlier literature~\cite{Ayral13,vanLoon14-2,Stepanov16-2,Terletska17}, as expected. Looking at $U<0$, $V>0$, the charge-ordered phases occurs at decreasing values of $V$ as $U$ decreases. The difference between the EDMFT and DB phase boundary stay approximately constant in the negative $U$ region.
At sufficiently large negative $U$, the checkerboard order appears even for $V=0$. By particle-hole symmetry, this corresponds to the diverging antiferromagnetic spin susceptibility at positive $U$ in DMFT. The present DB approach reduces to DMFT for $V=0$, so this was to be expected. However, this phase transition at $V=0$ violates the Mermin-Wagner theorem and is an artifact of the mean-field spirit of the approximation. This shows that this is a regime where nonlocal corrections over very long length scales should be crucial.

Then, turning our attention to $V<0$, we see that EDMFT produces a somewhat remarkable result. According to EDMFT, the phase separation occurs at exactly the same absolute value of $V$ as the checkerboard order, i.e. $V_c^{ps} = - V_c^{co}$. This is caused by the fact that the local properties of EDMFT are invariant (This assumes a certain amount of symmetry of the lattice, which is satisfied by the square lattice and hypercubic lattices in general, see \cite{Stepanov16} for more details.) under the transformation $V\rightarrow -V$. Since EDMFT takes momentum dependence into account on a very rough level, this invariance means that $\chi_{q} \rightarrow \chi_{q+(\pi,\pi)}$ under the same transformation, interchanging the checkerboard susceptibility and the compressibility. As a result, the divergence of these two susceptibilities occur at the same absolute value of $V$. The issue is that EDMFT largely decouples the kinetic and potential energy, only linking them up in the impurity model, and the impurity model does not know about the lattice structure. The $V\rightarrow -V$ symmetry should be broken by the \emph{interplay} of kinetic and potential energy, and EDMFT lacks the momentum structure of this interplay. 

This invariance of the EDMFT approximation is not a true symmetry of the  system, since repulsive and attractive interactions are physically quite different. Indeed, we find that DB gives different phase transition lines for repulsive and attractive $V$.

Unlike for repulsive $V$, however, the DB and EDMFT phase boundaries agree very well (but not exactly) for $V<0$ along the entire studied range of $U$. This shows that the phase separation and the checkerboard charge-density wave are physically different. The latter can be described accurately in a single-site, EDMFT, picture. From this, we conclude that nonlocal corrections to the compressibility are not very important. Earlier work on the compressibility in DB~\cite{vanLoon15}, also found relatively small corrections to the compressibility at half-filling.

\begin{figure}
\includegraphics{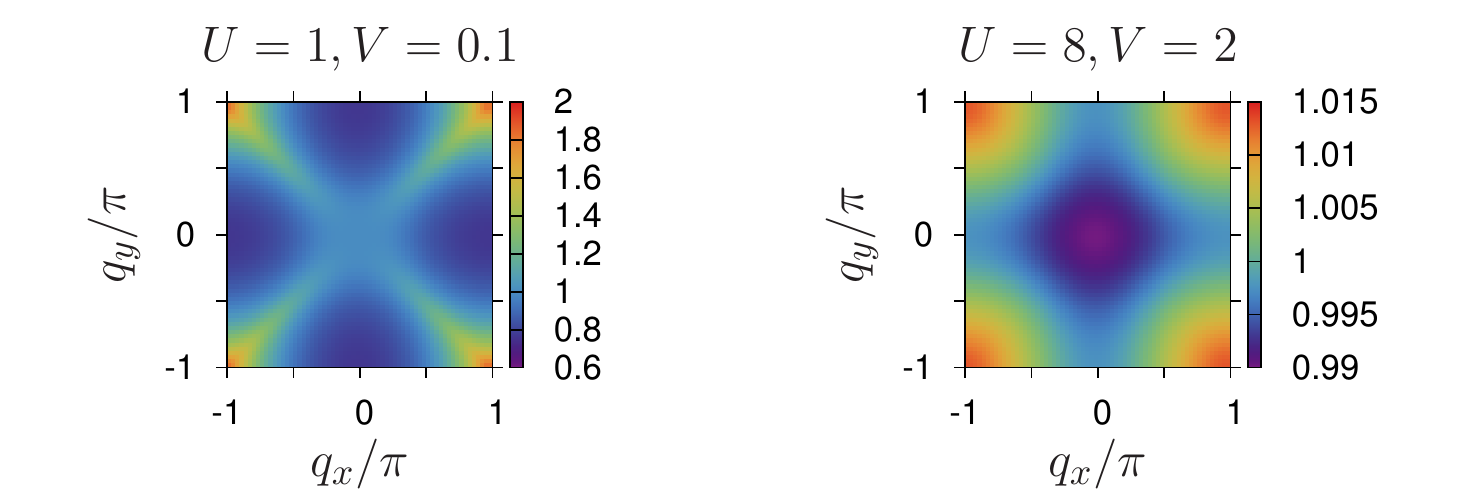}
\caption{Nonlocal susceptibility correction factor $1+\chi \tilde{\Pi}$ at $\omega=0$ as a function of momentum.
\label{fig:polarization}
}
\end{figure}

Figure~\ref{fig:pd} shows the dual correction factor, Eq.~\ref{eq:pi} across the Brillouin Zone. At small $U=1$, the nonlocal correction factor deviates strongly from unity, especially close to the checkerboard point $(\pi,\pi)$. In the center of the Brillouin Zone, the correction factor is approximately unity and corrections to EDMFT are small. This explains why the EDMFT and DB phase separation happens at approximately the same $V$. The nonlocal correction factor is actually smaller than unity around $(0,\pi)$, indicating that the nonlocal corrections decrease the susceptibility compared to EDMFT.
For larger $U=8$, only small deviations from unity occur since charge fluctuations are strongly suppressed by the on-site interaction. The momentum structure has changed, now the mimimum occurs at $(0,0)$ instead of at $(0,\pi)$. The small deviations from unity explain why the EDMFT and DB phase boundaries are so close at large $U$.

The $V\rightarrow -V$ symmetry of EDMFT does affect the DB calculations, since they use EDMFT as a starting point. In particular, all impurity quantities that enter the DB calculation are the same at $V$ and $-V$, and even a quantity like $\tilde{\Pi}$ is the same, since the nonlocal interaction $V$ does not enter it directly.
The dual self-energy $\tilde{\Sigma}$ depends on $\tilde{X}^{0}$ which does depend on $V$ explicitly, so this quantity is not symmetric under $V\rightarrow -V$. As a result, in DB calculations with inner self-consistency~\cite{vanLoon14-2} $\tilde{\Pi}$ will lose the $V\rightarrow -V$ symmetry after the first iteration.
An outer self-consistent DB calculation will also break this partial symmetry~\cite{Stepanov16}, since the impurity problem is no longer based on EDMFT. 

\section{Discussion}

Several schemes based on single-particle quantities only have been suggested to extend EDMFT~\cite{Sun02,Ayral12,Ayral13,Hansmann13,Stepanov16-2,Ayral17}. The advantage of these approaches compared to DB is that they do not require the (expensive) computation of the two-particle vertex of the impurity model. The approaches are mostly based on combining GW diagrams with a self-consistent impurity problem. 

There is discussion in the literature on how to best combine these ideas, and this ambiguity is especially important since the schemes give qualitatively different numerical results~\cite{Stepanov16-2}. Our findings here, and in particular the fact that the EDMFT and DB are very similar, indicate that the negative $V$ region could be a good test bed for these approaches: They are expected to interpolate between EDMFT and DB, adding only some nonlocal correlation effects. But here EDMFT and DB give near identical result, so we expect EDMFT+GW schemes to also give very similar results. Extensions of DMFT that give significant deviations should be under serious scrutiny for double counting issues. 

Checkerboard order becomes very favored for $U<0$. In fact, DB overestimates this tendency and shows a transition at $V=0$ already. This transition is related by particle-hole symmetry to a antiferromagnetic transition at $U>0$ which is forbidden by the Mermin-Wagner theorem. This makes this a very challenging regime even for single-shot DB calculations. The Mermin-Wagner theorem can be restored by enforcing appropriate local constraints~\cite{Vilk97,Krien17}.
This shows that the $U<0$, $V>0$ sector is an interesting place to compare computational approaches.

Phase separation can also occur in the Hubbard model without nonlocal interactions ($V=0$) away from half-filling~\cite{Otsuki14}. The situation away from half-filling differs strongly since there is no longer perfect nesting of the Fermi surface, and this reduces the tendency towards checkerboard order and antiferromagnetism so that other types of order become competitive.

\ack

E.G.C.P.v.L.  and M.I.K. acknowledge support from ERC Advanced Grant 338957 FEMTO/NANO.

\bibliography{main}

\end{document}